\documentclass[fleqn]{llncs}
\usepackage{isf}
\usepackage{version}

\title{Four Conceptions of Instruction Sequence Faults}
\author{Jan A. Bergstra}
\institute{Informatics Institute, Faculty of Science,
           University of Amsterdam, \\
           Science Park~904, 1098~XH Amsterdam, the Netherlands \\
           \email{J.A.Bergstra@uva.nl}}

\begin{document}

\maketitle

\begin{abstract}
The notion of an instruction sequence fault is considered from various perspectives. Four different  viewpoints on what constitutes a fault, or how to use the notion of a fault, are formulated. An integration of these views is proposed.
\end{abstract}

\section{Introduction}
\label{sect-introduction}
This paper aims at contemplating the notion of a program fault. Two aspects are considered together: how to use the notion of a program fault, a question which emerges even if making no commitment to the meaning of the phrase ``program fault''  is preferred, and what is a program fault, a question which emerges if making a commitment to the meaning of that notion is preferred.

Following the approach of \cite{Bergstra2012a} I will use a theory of instruction sequences serving as a model for a theory of programs. Consequently instruction sequence faults are taken as a model for  program faults. My arguments for adopting this approach are these:
\begin{enumerate}
\item A degree of freedom is obtained concerning the determination of meaning of terms and phrases which is unavailable when writing on program faults. If needed, the scope of claims concerning instruction sequence faults (or instruction sequence testing) can be limited by making the implicit assumption that instruction sequence theory is meant to refer specifically to a theory in the style of \cite{BL02a,BP04a,BM08,BM11a,BB2011a}.

\item Limiting the scope of claims concerning faults may be helpful to avoid making assertions about instruction sequence faults that unnecessarily contradict existing literature on program faults.
If conclusions concerning instruction sequence faults have convincing arguments that transfer to the case of program faults, that contradiction with existing literature is only present for readers who indeed, have been convinced by the entire chain of arguments, including the inductive step from instruction sequences, narrowly understood, to programs at large. That inductive step cannot be taken for granted in any general sense, and its validity needs to be checked for each individual kind of assertion about instruction sequences that makes sense for programs in general as well.
\item 
The notion of a computer program not only lacks any common definition, definitions of the concept of a computer program that have not gained wide acceptance are also hard to find. For instruction sequences matters of definition have been settled in \cite{BL02a,BP04a,BM08,BB2011a}. Defining program faults is made much harder in the absence of a definition of program.  These difficulties are present to a lesser extent when considering instruction sequences.
\item 
Specialized notions such as instruction sequence execution can be provided with comprehensible definitions (see \cite{Bergstra2011c,Bergstra2012a}), something which seems to be impossible for the more general notions concerning programs.
\item 
The use of new terminology is less provocative when writing on instruction sequences than it is when dealing with programs. For instance, following \cite{Bergstra2011c} I will write that an instruction sequence is put into effect in cases 
where it would be customary for work on testing to write that a program is executed.%
\footnote{Putting into effect is more general because it includes execution (if the instruction sequence is considered executable),  interpretation, compilation followed by execution, compilation followed by interpretation, interpretation combined with 
just-in-time compilation, compilation followed by interpretation combined with just-in-time compilation, compilation involving significant optimization steps followed by execution, and so on, at the most liberal end including forms of simulation.} 
The putting into effect of an instruction sequence will also be referred to as an effectuation.%
\footnote{The range from execution to simulation concerns a mechanical classification of instruction sequence 
effectuations, by describing different mechanisms that are used to get it done. In Section \ref{TCoISE} I will provide a teleological classification of effectuations, by describing different motives why it is done.
In \cite{AdrionBC1982} the phrase  conceptual execution is used if a manual process for performing the steps prescribed by an operational semantics is available. In the terminology that I prefer, ``conceptual execution'' is avoided and is replaced by ``manually putting in to effect'' or ``manual effectuation'', which can  be considered  an instance of simulation.}
\end{enumerate}

The approach of analyzing instruction sequences as a model for analyzing programs has disadvantages too. I mention the following issues, without any claim of completeness:
\begin{enumerate}
\item 
There is no industrial scale practice of instruction sequence engineering and usage. All notions that make essential use of the industrial scale of certain processes can only be mimicked in the world of instruction sequences. That is most apparent with the concept of an instruction sequence effectuation failure. Such a failure is unlikely to cause the crash of a spacecraft, and is much more likely to be limited to a mismatch between expectation and observation in experimental and properly controlled circumstances where losses are predictably low.
\item 
Due to the non-existence of a proper instruction sequencing practice there are no experts on instruction sequencing, and no communities providing standards for products and processes. Below I will use the notion of an ``instruction sequence product expert'' and similar notions. These must be read in a hypothetical fashion, that is the reader is asked to imagine another world in which such experts exist.%
\footnote{The clarity which is gained by instruction sequence having an unambiguous definition, is lost in part by the fact that related roles of human engineers and users are hypothetical.}
\item 
No form of generalization of results concerning instruction sequences to results on programs can be taken for granted.
\item 
Programming is a much richer world than instruction sequencing and conclusions drawn concerning instruction sequences may only extend to aspects of programming of lesser importance.
\item
It may be held against  investigating instruction sequence faults (as a model of program faults) that instruction sequences (in the technical sense meant in this paper) are so simple in comparison to program notations used in practice that there is little difference between having an intuitive grasp of their (intended) operational meaning and having available a full formalization of their operational semantics together with a correctness proof system. Having the latter ingredients available  opens the door to a formal definition of fault in the style of Laski's \cite{Laski1997},%
\footnote{It seems that Laski's analysis of faults admits many variations, I will use his analysis in this paper as representing an approach where fault is supposed to have a theoretical definition in principle (though perhaps not in practice) independent from any effectuation.}
in a way which may be unavailable for industrial strength program notations.%
\footnote{That simplicity  is an illusion, however, because already with a fairly limited set of instructions (e.g. the instructions proposed in \cite{BL02a}) a combinatorial explosion of operational options emerges, each of which comply with the intended meaning of the catalogue of  instructions at hand. A typical example is what to do when a jump moves outside the given range of instructions, that is the program counter either will become too low or too high. It may be treated as a deadlock (incorrect termination without warning), as a live-lock (endlessly ongoing processing), as an error (incorrect termination with warning), as an instance of correct termination, as an idle step, or it may be excluded assuming that static type checking precedes any run and the detection of syntactical faults will prevent the instruction sequence from being put into effect. Excess at both ends of the instruction sequence might be handled differently and in this way some 30 different options for an operational meaning arise. With other instructions providing similar degrees of freedom thousands of options are easily generated. Each of these options can be implemented and such implementations are all somehow reasonable, because the differences appear only in marginal cases. However, each specific operational meaning gives rise to its own 
specific concept of fault. A user may have in mind that a machine on which an instruction sequence is put into effect 
realizes exactly one of these semantic options, and, not knowing which one, the user may either assume a probability distribution over different semantic models (and correlated realizations on a machine), or the user may run tests in order to find out which semantic model has been the inspirational source for the machine architects. In either case simply basing a concept of fault on a theoretical model of machine behavior and instruction sequence processing is implausible.}

\end{enumerate}

Because I hold that the mentioned advantages outweigh the listed disadvantages, I  am confident that investigating instruction sequences faults is worth the time. Obviously there would be no issue regarding instruction sequence faults  if program fault were an unproblematic concept. But on the contrary the notion of a program fault is somehow mysterious, 
and below in Paragraph \ref{what-is-the-problem} I will list some of the questions one may contemplate concerning program faults.
\subsection{Restriction to non-reactive and sequentially computed functionalities}
I will assume that instruction sequences are used to express non-interactive, non-reactive functionalities which are computed in a sequential fashion, that is partial operations on a state, fully specified by a single state transition without taking any notice of what takes place in between. This constitutes a significant simplification because instruction sequence notations are meant to be more general. Nevertheless it fits well with most conventional work on program testing where it is supposed to be more or less obvious when the result of a computation complies with the information that an oracle provides, the oracle being simply a partial function on states, or more generally a binary relation on states.

Testing reactive functionalities is not an obvious matter, and definitions of compliance require much care. In \cite{Tretmans1996} the foundations were laid for a theory of testing for reactive and interactive systems.%
\footnote{In \cite{EdelsteinFGNRU2003} a desciption of testing of multi-threaded programs is given, which still fits
the terminology and concepts known from sequential systems.}
For my current objectives in this paper
the mathematical properties of compliance definitions for reactive systems seem not to enter the picture, 
because I hold that
the main conclusion, namely that the notion of fault depends on the notion of test, and that it is independent of both formalized correctness proof and formalized operational semantics, will not change by taking reactivity into account. 

I do not claim that the assertion, that ``faults in instruction sequences expressing sequential algorithms for 
non-reactive systems must be defined on the basis of a conception of testing'', transfers to a setting with reactive systems. 
It may be the case that compliance definitions for reactive systems are less based on straightforward operational
intuition to such an extent that there can be no understanding of those notions without preparatory formalization. Still that
state of affairs is consistent with a dependency of fault on test in the non-reactive, sequential case.%
\footnote{The merits of testing are not limited to software quality management viewed from the perspective of software failure prevention. In \cite{Colburn1991} a wider perspective is sketched that allows an essential and probably permanent role for software testing outside the arena of software failure prevention.}

\subsection{What is the problem?}
\label{what-is-the-problem}
In an arbitrary order I will now list some of the questions that arise when contemplating  program faults.
\begin{enumerate}
\item 
A clear indication that program fault is a problematic concept is to contrast the massive occurrence of  the term ``fault'' in the software engineering literature, most notably in the literature on software testing, with the almost universal silence about the meaning of that term. Explanations are mostly limited to stating the contrast between fault (said to be a static concept) and failure (usually referred to as a dynamic concept), often in a context also distinguishing mistake and error.%
\footnote{For the notions of mistake, fault, error, and failure I refer to \cite{Bergstra2012a,Laski1997,DaranT1996}. There is little uniformity in the use of these terms in the literature on software testing, but there seems to be a growing consensus that programs cannot contain errors, because that is a dynamic concept, nor failures for the same reason. Programs don't contain mistakes either, because mistake is a dynamic concept concerning programmer behavior. Programs may contain faults, however. Faults are caused by mistakes, and cause errors, some of which qualify as failures. Both notions of causation require a possible worlds analysis substantiating the hypothetical existence of a ``better world'' in which the fault is absent, most other aspects being the same. In \cite{Hatton1998} the term error is considered ``a problematic term used in different ways in different standards''.}
In that majority view program faults
are fragments of programs which can be understood to cause the occurrence of errors or failures when a program is being put into effect.
\item
In \cite{MunsonNS2005} the remark is made that
 ``unfortunately there is no particular definition of what a software fault is''.
 In addition it is claimed in \cite{MunsonNS2005} that a definition of a software fault must make that notion quantifiable, that is both the number of faults in a program and the size of each fault must admit objective measurement. This plausible  requirement is hardly ever taken into account in writings that make use of the concept of a  program fault.
\item 
The idea that program correctness is about the absence of faults (so-called bugfree programs)  is misleading. Program correctness is a holistic concept and an engineer's inability to prove, or to provide evidence for, the correctness of a program is quite remote from the ability to spot a  fault in the same program. What program fault and program correctness have in common is the remarkable absence in both phrases of requirements and specifications, suggesting that a program as such could ever be correct, incorrect, fault-free, or faulty.%
\footnote{So-called self-documenting programs may be such that their correctness can be asserted with out taking additional specifications or requirements into account. There is no indication, however, that programs are commonly understood to be self-documenting.}
\item 
There is a remarkable richness of terms for indicating properties (qualities) of software that are generally expected to correlate with the absence of faults or more precisely the low frequency of fault occurrence: software may be reliable (the opposite of containing many faults), dependable (extending reliability with a positive judgement about requirements), compliant (with specifications), correct (relative to a specification), being of high quality, and simply usable. In any case, the precise relation between these terms is not easy to grasp, and for instance one may ask whether or not an identification of ``containing few faults'' with ``being reliable'' is reasonable.%
\footnote{Reliability may also be understood to pertain to the absence of failures rather than to the absence of faults.
Is software reliability the guaranteed absence or at least low probability of software failures? That seems to be an attractive position, though it makes instruction sequence reliability a less credible notion for the same reasons that instruction sequence failure is somewhat less credible. This interpretation of software reliability would confirm all positions on the assessment of software reliability endorsed in \cite{Fetzer1991}. Instruction sequence reliability is then understood, for
a given instruction sequence, as the improbability of instruction sequence effectuation failure. Obviously the question how to determine such a probability in a specific case is far from trivial.}

\item
An obvious idea is that a program fault is a program fragment which constitutes an obstacle to achieving certain goals. That immediately leaves one with at least the the following options: proof obstacle (frustrates a proof), compliance obstacle (causes an effectuation failure), software product expert approval obstacle (causes an expert to refuse issuing an approval of the program), software process expert approval obstacle (uncertainty about faults, for instance due to inadequate testing,  may cause a software process expert to approve of a specific software process).
Now the following question emerges: is this a rudimentary classification of faults, or are these mutually inconsistent views on faults from which one may choose, or should only compliance obstacles be considered faults, with the three other views being rejected.
\item 
\label{located-faults}
Assuming that viewing a fault as a program fragment which may (and under some conditions will) cause a effectuation failure is considered plausible, it must still be admitted that the phrase program fragment is vague. Using instruction sequences instead of programs a fragment of $x$ may be understood as a subsequence of instructions of $x$. Now a simple problem emerges: what lengths are admitted for a fragment $f$ of $x$ which is considered faulty. If $x$  has length $k$ then it may be considered unreasonable if the fault has the same number of instructions, in fact one expects that the length $l_f$ of $f$ must be small in comparison to $k$. But  what is considered small is quite arbitrary and it may vary from context to context.
It follows that behind a notion of a fault as a part of a program lies some notion of being small, which needs to be made explicit, for instance by requiring that $l_f < k . C$ for some constant $C$. In the paper I will assume $C = 0.05$. Another complication is that a fault may be distributed over various locations. A $n$-located fault in $x$ is a family $f^n = (f_1, ..., f_n)$ of disjoint subsequences of $x$ which may (and under some conditions will) constitute the cause of an effectuation failure of $x$. The length $l(f^n)$ of an $n$-located fault is the sum of the lengths of its parts. Again a requirement like $l(f^n) < k . C$ will be needed.  
\item
Work aiming at defining program faults seems to be unconvincing. To substantiate that claim I will briefly discuss an important example of such work.
A program which is investigated for containing faults will be called a program under fault assessment (PuFA).
In Laski \cite{Laski1997} the notion of a program fault has been analyzed in depth.   The following conclusions were 
reached in \cite{Laski1997}: 
\begin{enumerate}
\item what can be adequately defined is the notion of a software component being faulty (which is identified with it containing a fault),  
\item that definition requires the availability of a sound and complete proof system for proving program correctness (compliance with the specification) from component specifications, 
\item the definition (of a component containing a fault) also presupposes that a formal operational semantics for the used program notation is given, and finally
\item the definition (of a component containing a fault) depends on combined use of these ingredients
for both the PuFA and for a modification of the PuFA in which precisely one of its components has been changed.
\end{enumerate}

In spite of the clear and comprehensive account given in \cite{Laski1997}, which is the best reference I could find on the question ``what is a (program) fault'', I do not agree with the views expressed in that paper. In particular I cannot 
believe that the notion of a program fault 
(for programs written in a certain program notation) needs to  be built on top of the notion of a proof system for program correctness (for that same program notation). This dependency of fault on proof seems to be quite counterintuitive. I also
do not believe that the notion of a program fault depends on the availability of an operational semantics  (or of any theory capable of predicting what will happen when a program is being put into effect%
\footnote{In \cite{Bergstra2011c} I have explained why I prefer the phrase ``putting into effect'' over the term ``executing'' in the case of instruction sequences, and the same preference extends to many other program notations for similar reasons.}) 
either, although that suggestion is definitely less counterintuitive.%
\footnote{Indeed the vast majority of papers on program testing, including the papers referred to in \cite{Middelburg2010b} and \cite{Bergstra2012a}, do not mention either program verification or formal operational semantics in more than a casual way, often suggesting that such ``formal methods''  can't cope with practical matters, from which the inference is made that testing is a programmer's best option for the time being.}

\item 
The relationship between faults and testing is not obvious. At first sight it appears that a theory of instruction sequence testing needs to be based on a theory of instruction sequence faults but I will conclude below that the connection between testing and faults cannot be that simple.

Many papers on program testing claim that program testing is a methodology for discovering the existence (and perhaps finding the location) of program faults.%
\footnote{A brief history of the concepts of program testing has been given in \cite{Middelburg2010b}. This paper highlights two remarkable open questions concerning program testing: how to include the experimental aspect in the definition of program testing, and how to find theoretical evidence for  its relative usefulness in comparison to other methods of program quality engineering. This relative usefulness seems to be confirmed by the overwhelming attention paid to testing in comparison to program mathematics and comparative static program analysis. Nevertheless, to have a more direct insight in this matter will be interesting and useful.}
In \cite{Bergstra2012a} I have made an attempt to understand the concept of program testing (using the thought experiment of the plan to work out a theory of instruction sequence testing). Different perspectives on program testing  were surveyed in some detail. It turns out that a variety of views exists on the essence of program testing, while a majority view seems to be that program testing can be characterized by the combination of its experimental method and the objective to combat or to control the problems caused by program faults.%
 \footnote{The phrase ``program mistake'' is often used (for instance in \cite{LokazyukG2005}), in particular in general news items explaining that software is to be blamed for some problem (e.g. recent news items concerning causes of problems with the Phobos Grunt spacecraft.) I prefer to speak of a programming mistake, that may have caused one or more  program faults.}
This majority view is questionable,%
\footnote{In \cite{MaesDeVries08} an exposition is given on how to deal with self-reinforcing majority views that one disagrees with.  It is argued that empirical work is insufficient for that task and that theoretical work is needed. I agree in practice, with that viewpoint, though in the case of program faults empirical work might well bring additional insights. For instance in \cite{MunsonNS2005} it is stated, ``on the basis of their careful examination of program faults over the years'',  that ``the overwhelming number of faults that are recorded as code faults are really design faults''. Now the consequences of design faults are definitely not caused by programmer mistakes. Moreover if
system behavior conforms to the design objectives there is no program fault present (though there may be a software fault present in case the requirements have been laid down in a document which is considered to belong to the software).} 
in spite of its seemingly compelling presentation in many papers,
however, because of the hypothetical nature of the effectuation that is used to explain the causal relation.%
\footnote{Typically, instead of explaining what a program fault is, it will be stated how much harm to society is inflicted by the frequent occurrence of program faults. That leaves readers (and authors) without doubt: what causes so much harm must exist. But such arguments might be flawed: ill-conceived program designs, based on mistaken requirements, may inflict similar or even greater harm and may do so especially after having been implemented without fault.}
\item
When contemplating faults as causes of failures  the teleology of effectuation cannot be ignored. 
If  we assume that in order to prove (meant informally) that a program fragment is causing an error or a failure when the program is being put into effect, the program needs to be actually put into effect on suitable inputs in a suitable context, the question arises what qualification this effectuation might carry. I will classify such effectuations below in Paragraph \ref{TCoISE}, suggesting that one may choose between test and use, with demonstration constituting one of two kinds of use. If the effectuation is classified as a test, the notion of a fault is dependent on a test, and a remarkable circularity arises.%
\footnote{The circularity being that tests are needed to bring about the concept of a program fault in the first place. More specifically, given a PuFA, tests are needed (i) to demonstrate the occurrence of an error or failure when the program is put into effect, and (ii) to demonstrate that by appropriately modifying a fragment claimed to be faulty, the occurrence of that same error or  failure can be avoided, thus establishing the key characteristic of a fault: that it can be provisionally repaired (and for that reason that it might just as well not have been there in the first place). The reparation is termed provisional because it may be unknown to what extent the modification of the fragment claimed to be faulty gives rise to the occurrence of new failures for other inputs or in other contexts. Regression testing is commonly referred to as the experimental activity which investigates the claim that the provisional repair is a step forward indeed. Fault injection, see e.g. \cite{EichingerBH2008} provides crystal clear instances of program faults, as far as repair potential is concerned. However, that an injection indeed creates a fault needs to be established by demonstrating a resulting failure. The techniques of \cite{EichingerBH2008}, making use of call graph statistics, may highlight injections as if they were faults even if no failure is caused by them. For that reason the experimental set-up of \cite{EichingerBH2008}   restricts attention to cases where failures are indeed present.}
If it is an instance of ``use'', the concept of program  fault becomes dependent of the concept of program use, which is rather implausible.

I hold that a demonstration is not about finding new information but about transferring known information to one or more other agents. This eliminates demonstration as a qualification for putting a program into effect with the objective of proving that a program fragment is faulty.%
\footnote{Although being a fault seems to require a definition based on testing, detection of faults need not follow that route. For instance in \cite{DeeprasertkulBO2005} a data-base of patterns of plausible constructs is used to match program fragments before compilation. It is claimed that this form of filtering performs better than dynamic experimentation would do. Is seems that this technique does not guarantee that a program fragment which was found suspicious will 
cause a failure.}

The remaining option left open is to qualify an effectuation as a  test, and to develop a viewpoint from which the circularity just  mentioned need not be considered a prohibitive obstacle. In \cite{Bergstra2012a} I have not made an attempt to define instruction sequence testing.%
\footnote{In \cite{AvitalTeni08} the implicit suggestion is made that testing is no more than simulation. As such it plays an important role in design.  If this were true one can do away with the term testing, an interesting perspective. This very liberal viewpoint seems to fall short in terms of making explicit the specific objectives of testing. In \cite{AmmannOffutt2008} a definition of testing is given which pays too little attention to the experimental aspects of it, the same may be said concerning 
\cite{Beizer1990}. A better definition must be somewhere in between of these two viewpoints. Whatever definition is given, it should preferably follow the guidelines of \cite{AvitalLBBD2006} and incorporate a so-called positive lens.}
The curious relation between fault and test implies that the mentioned circularity needs to be tamed by means of an appropriate definition of testing. Working towards a definition of instruction sequence testing is not the purpose of this paper either, but it can be concluded that a definition of the form ``testing is executing an instruction sequence with the intention to detect faults'' cannot work for the simple reason that such a definition excludes the application of testing for 
characterization objectives concerning faults.%
\footnote{In \cite{Bertolino2004} the existence of a gap between theoretical work on testing and industrial practice is commented. An increased focus on comparative empirical research on fault detection methods  
is put forward as a remedy.
Besides that, it is suggested that testing research ought to pay attention to all aspects of testing rather than mere test case generation which had dominated the scene according to \cite{Bertolino2004}. This suggestion, however, ultimately depends on having a reliable  definition of program testing at hand, which as far as I can see still poses a challenge.}

\end{enumerate}

These arguments in support of the viewpoint that the concept of a program fault is non-trivial are quite varied and the given listing may admit  a better structure. But the listing suffices to justify a further investigation of program faults, and in my view the issues raised in the listing also justify doing so via the detour of investigating instruction sequence faults.

\subsection{Structure of the paper}
After a survey of advantages and disadvantages of instruction sequence theory as a means for investigating topics on programs and programming, a detailed investigation of so-called mechanical faults is made. Then so-called instruction sequence software defects are discussed, together with the notion of a persistent fault. Subsequently the most common viewpoint on software faults, that is the mechanical view, is complemented with three additional views on faults. The mechanical view on faults is split in two extremes, a logical view and  and an empirical or test based view. Between these views are many intermediate options. 

The intended audience of the paper consists of readers with an interest in the theory an application of instruction sequences as well as of readers having an interest in the theory and practice of program faults.

\section{Mechanical instruction sequence faults}
The  concept of a program failure seems to be prior that of a fault in the sense that failure can be defined without making reference to fault. That observation suggests contemplating the following three phrases first: ``software failure'',%
\footnote{In \cite{Alzamil2008} a failure is called an ``incorrect behavior''.}
 ``program failure'',  and  ``instruction sequence failure''. I will assume that these phrases are a shorthand for respectively 
``software effectuation failure'', ``program effectuation failure'', and  ``instruction sequence effectuation failure''.

An instruction sequence effectuation failure is a deviation or discrepancy between the behavior resulting from effectuating an instruction sequence and the specification of the behavior intended to be implemented by that same instruction sequence. This discrepancy can become manifest as a theoretical fact: the formal operational semantics of an instruction sequence does not satisfy a formally specified intended requirement. Alternatively that discrepancy may show up in an 
experimental fashion, that is through a test, or during usage.%
\footnote{Asserting that an instruction sequence effectuation  failure is a systems failure caused by an instruction sequence fault is plausible at first sight. At closer inspection that definition presupposes that an instruction sequence plays the role of a system component causing a failure during an activity that involves an effectuation of the instruction sequence.
What complicates this view is that the system involved  makes use of a component (called  instruction sequence) that serves as a model of a program rather than as the software itself. As a consequence the entire system automatically acquires model status as well, and the failure reduces to a mere event, without the potential of inflicting damage.}

The notion of a program effectuation failure can be understood like the instruction sequence effectuation failure just mentioned, but ``software effectuation failure'' refers to a different notion  because it may also bring about a problem with the specification, which constitutes a part of the software. In other words: not every software effectuation failure involves a program effectuation failure.%
\footnote{The notion of a software failure can be found for instance in \cite{MockusW2000}. There I understand that a software failure is meant to be a system failure caused by a software fault. This view introduces a dependency of failure on fault which I prefer to avoid. 
The causes of a system failure may be quite hard to assess, even if  there is no doubt that software is critically involved. For a remarkable survey of diverging diagnostic accounts concerning the same failure see \cite{Nuseibeh1997}. That paper suggests that what is a software failure at first sight may be considered a consequence of inadequate risk management at closer inspection.}
Although technically redundant because it is merely a shorthand as just indicated,  the phrase ``software failure'' has an appeal and creates a sense of urgency which the phrases ``software fault'' and ``software defect''  do not entirely match.

I this section I will describe what I consider to be the most plausible link between instruction sequence effectuation failures and instruction sequence faults. This link implies a mechanical view on faults. Later in the paper that view is contrasted with several other views. Moreover it will be outlined that a mechanical view can be based on empirical notions as well as on formal and theoretical notions. There is in fact no single mechanical view because a single assessment may have several observations as inputs and each of these observations can in principle be replaced by a proof of a mathematical fact, however complex that asserting that fact may be and however hard to obtain that proof may be.

\subsection{Teleological classification of instruction sequence effectuations}
\label{TCoISE}
For an understanding of mechanical faults it is useful to avail of a teleological classification of instruction sequence effectuations. Such a classification provides different  indications on why an effectuation is performed. The teleological classification must not be confused with a mechanical classification of effectuations, which classifies the various ways in which an effectuation may take place. Here is a proposal comprising only two forms of effectuation: test and use. Both forms are subdivided in two cases, thus yielding four categories of effectuation.

\begin{description}
\item{\em Confirmation test.} This is an effectuation performed to prove (not in a mathematical sense) a fact about the instruction sequence, for instance that it delivers a certain output  (or  an output that satisfies a given condition) on 
a certain input.%
\footnote{An oracle, see e.g. \cite{White1987}, may be used to check whether an outcome is confirming.}
\item{\em Experimentation test.} This is an effectuation performed to determine whether or not the result complies with certain criteria. It produces new information, rather than to confirm a hypothesis.%
\footnote{This class of effectuations comprises so called alpha tests and beta tests.}
\item{\em Demonstrational use.} A demonstration is an effectuation that is performed by an agent (the demonstrating agent) in order to show other agents information about the instruction sequence which is known to the demonstrating agent already. (It may either be a demonstration of a failure or of a success, in both cases the demonstrator knows in advance what will happen. In the absence of that information the demonstrator performs an experimentation test, perhaps without admitting this to the audience.)
\item{\em Practical use.} An effectuation that serves the effectuator's  objectives in the absence of any gain or loss of information concerning the instruction sequence represents practical use. I consider it to be a common understanding that 
practical use, or the perspective thereof, is the primary reason for the instruction sequence's existence.%
\footnote{Here the approach to model program theory by instruction sequence theory becomes less plausible, because few, if any, instruction sequences reach the maturity that one may speak of intended use. Toy instruction sequences have demonstrational use at best.}
\end{description}

\subsection{Unqualified instruction sequence effectuation cannot fail.}
If instruction sequence $x$ is effectuated and  a discrepancy between the behavior generated by effectuation and the specified behavior is observed, common parlance on testing suggests that a (real) failure is observed. I hold that this is unsatisfactory terminology, and that unqualified instruction sequence effectuation cannot fail. One needs to know in addition what kind of effectuation has taken place. If it is a confirmation test the confirmation has failed. That in itself is not an indication of the existence of any fault, on the contrary, it may defeat an attempt to locate a fault.

If an experimentation test, a demonstrational use, or a practical use,  was performed, in technical terms a failure was observed. But these three cases differ significantly. In the case of an experimentation test finding the mismatch is a success, the term failure being a mere label of an observation. In the case of a demonstration the event of observing the failure is perhaps frustrating and unhelpful making the demonstrating agent insecure if not angry (if she/he has made an investment in the process of demonstration),%
\footnote{If the demonstrator has noticed the failure but the demonstree did not notice the failure that very strike of good luck may be considered a success by some demonstrators.}
but in the case of practical use it may have grave consequences. Only in the case of practical use, one may speak of a real failure, in the case of a demonstration it is the marketing which fails, not the effectuation (because for the demonstree it may constitute a revealing success, depending on the demonstree's degrees of freedom concerning the matter), and it may reveal a lack of product knowledge at the demonstrating agent's side. In the case of an exploration test the term failure is a mere technical term from program testing theory which has nothing to do with any real failure and  in that case a  failure is most likely taken to be a success.

\subsection{Confirmation tests in more detail}
I will provide a definition of a confirmation test which suffices for this paper. This definition has an preliminary status and it plays an auxiliary role for this paper only. The definition  is not meant to serve as a definitive definition of test, or of a particular kind of test.

A test case is a triple $(d,U,k)$ with $d$ input data (test case input), $U$ a set of possible output data, and $k$ a natural number indicating that an effectuation should reach at least those states of a computation which are within $k$ steps of the operational semantics of $x$ on $d$.%
\footnote{It must be stressed that a test case is independent of the instruction sequence for which it may be used. Indeed it is a known programming method to start with the generation of test cases before construction a program that will pass the corresponding tests.
For the details on the operational semantics of instruction sequence effectuation as well as the counting of steps 
one may consult \cite{BM12a}.} The test case
contains both oracle information (membership of $U$) and timing information ($k$) concerning a (confirmation) test. 

A confirmation test (of an instruction sequence $x$) is an effectuation (of $x$) initiated and monitored from the perspective of a particular test case, say $(d,U,k)$. The test succeeds (complies with $(d,U,k)$) if the effectuation of $x$ on $d$ terminates and the resulting state is in $U$, otherwise it fails, where it is assumed that at least $k$ steps of the operational semantics of $x$ have been taken into account. For a successful confirmation test it is permitted that it covers more steps than the initial $k$ steps of the formal operational semantics. Thus the test fails if either it is stopped by the test execution mechanism having performed at least enough steps to ``see'' what happens within the  initial $k$ steps of the formal operational semantics, or if it terminates in a state outside $U$. 

If $U$ is large or infinite another instruction sequence may be needed  to check automatically whether or not a confirmation test has succeeded. Effectuating $x$ on $d$ in such a way that enough steps are taken, and subsequently stopping the computation followed by an assessment of output (if any) is contained in $U$ will be referred to as executing the test as specified by the test case at hand. Executing a confirmation test for an instruction sequence need not involve executing the instruction sequence, it suffices that some (form of) effectuation takes place assuming that enough steps are taken into account.

\subsection{Defining faults in terms of effectuation failure causation}
The application of the teleological  classification of effectuations to the case of instruction sequence faults is as follows: in order to assert the existence of a fault $f$ in instruction sequence
$x$ causing failure $F$ on input $i$, the following state of affairs is necessary and sufficient:
\begin{itemize}
\item 
The fragment $f$ is in the simplest case a substring of $x$ (a 1-located fault in the terminology of \ref{what-is-the-problem} 
nr. \ref{located-faults}) and in more complex cases a family of 
substrings of  $x$ ($n$-located faults with $n > 1$).  The length (or sum of lengths) of the fragment(s) must be bounded by say 5\% of the entire instruction sequence length.%
\footnote{Different faults must be non-overlapping fragments of an instruction sequence. Together faults may not exceed say 25\% of the instruction sequence. These requirements enable the counting of faults in $x$. Besides a syntactic size a fault also has an impact, that is a degree of caused deviation from intended behavior upon effectuation. Measuring impact is difficult, but in order to allow distinguishing errors it must be assumed that the impact of combined faults exceeds the impact of individual faults.}

\item
Some activity, for instance an experimentation test or an inspection has led to the detection of $F$ (with an effectuation of $x$ on input $i$), and some form of subsequent reasoning has led to the proposal to spot $f$  as the candidate fault.
\item
A candidate repair $f^{\prime}$ of $f$ is proposed, which can be applied by substituting the $f$ by $f^{\prime}$ in $x$, thus turning $x$ into $x^{\prime}$ which remedies $F$. 
\item
That a repair is in fact obtained is established by means of the repair confirmation for $f^{\prime}$: a confirmation test of $x^{\prime}$ on input $i$ that avoids failure $F$ (and any other failure).
\item 
Assuming that a number of tests has already been applied to $f$ each leading to a valid result (asserting that requires the availability of an oracle), these tests are repeated for $x^{\prime}$ (regression testing) to secure that no new failures have been introduced by fixing $x$.%
\footnote{Regression testing only provides some evidence that no new failures result from fixing $x$ by means of the candidate repair. Nevertheless, the application of regression testing seems to be essential for the definition of a fault, and more remarkably, its role makes the concept of a fault technically dependent of the stage of testing and debugging of an instruction sequence, because the later in the process the fault is spotted the more effectuations for regression testing will 
be and must be performed.}
\item 
From so-called non-functional requirements bounds $k$ for the confirmation test cases involved must be derived.%
\footnote{Confirmation testing in the absence of non-functional requirements, seems to be impossible in principle.}
\item 
(Optional: minimality constraint.) For $f$ to be recognized as a fault that causes $F$ it may in addition be required that no proper substring $g$  of $f$, together with a modification $g^{\prime}$ of $g$ could have been taken as a candidate fault for repairing $x$ instead. 
This additional requirement adds much complexity to the concept and it seems to be of little practical value.
\item(Optional: happy end constraint.)  The happy end constraint requires that (i) a series of (perhaps multi-located) faults can be found, and (ii) a series of corresponding repairs can be found, so that, (iii) if these repairs  are applied in succession a correct instruction sequence results, and (iv) the sum of the sizes of the respective faults does not exceed a given threshold.%
\footnote{The existence of a series of candidate faults and candidate repairs which when applied in combination transforms the instruction sequence into a correct one must be distinguished from the capacity of some method for  fault finding and debugging to determine such a series. There may not be a method for doing so even if the series exists.}
 \end{itemize}
The role of regression testing in the definition of fault  is not obvious. The following remarks can be made:
\begin{itemize}
\item One may imagine a context where a complete collection of test cases is given and the ``idealized regression criterion'' on a repair is formulated as follows: at least one confirmation test that did not succeed before the repair succeeds thereafter, and all successful confirmation tests remain successful, where the latter implication and quantification are meant in a formal manner. That means that ``a successful test'' refers to a test that might have been executed and for which the theory predicts that it then would have been successful.
\item Regression testing approximates the idealized regression criterion from below while admitting false positives. 
A false positive may 
occur if $f$ is considered a fault because regression testing has failed to  uncover a test case where replacing $f$ by $f^{\prime}$ does not succeed in spite of the fact that this replacement does introduce new failures (i.e. that such test cases exist outside the collection of test cases used in regression testing at that stage in the process). 
\item 
From a logical point of view these matters are quite complicated and there seems to be an unavoidable discrepancy between an ideal mathematical notion of fault and a pragmatic one. If one aims at giving a logical definition of fault taking ``avoidable cause of failure'' as the point of departure, the need to provide a candidate repair exists as well as in the case making use of confirmation tests. Then it is plausible to require of a reparation  that after its application no new failures are introduced (idealized regression criterion). In a formal world providing the latter fact requires a formal verification which must take two instruction sequences into account. That verification is likely to be of a comparable complexity to proving correctness of the original instruction sequence with respect to its specification. This leads to the following conclusion: (i) having a formal definition of a fault in mind one cannot maintain that testing suffices for finding faults, because testing is probably insufficient to provide conclusive evidence that no new failures have been introduced by the proposed repair, and (ii) having a confirmation test based definition of fault in mind one cannot guarantee that repairing a fault is a step forward that need never be undone, (iii) having a confirmation test based definition of fault in mind false positives cannot be avoided in practice due to the limitations of regression testing.%
\footnote{I conclude by taking these observations together that the slogan that ``testing allows to prove the existence of faults but not their absence'' is wrong. The slogan ``testing allows to prove the existence of failures but not their absence''  is valid, however.}
\item
Without making use of regression testing proposing a notion of fault that admits an effective characterization seems to be impossible. Thus the notion of a fault is not only based on confirmation testing but more generally it is based on regression testing.
\end{itemize}

\subsection{Usage of mechanical faults: transformational improvement engineering}
Faults are used in the following instruction sequence improvement process, which I will refer to as transformational improvement:
\begin{enumerate}
\item 
look for failures via experimentation tests,
\item
infer candidate faults from these failures and,
\item 
find candidate repairs so that confirmation tests (by way of regression testing) allow to categorize the candidate fault as a fault, and
\item
categorize its candidate repair as a repair, which,
\item 
is subsequently applied to the instruction sequence at hand. 
\end{enumerate}

This improvement process will not escape from backtracking because dead ends cannot always be avoided, at least not in practice. Even if a fault is defined via confirmation tests as above it is not entirely obvious that repairing a fault, once found, is advisable. Indeed, that only works out well if the happy end constraint is satisfied and if the software improvement process is able to find a series of faults/repairs that finally leads to a correct instruction sequence.

If the happy end constraint is included the definition of a fault, asserting that a fragment of an instruction sequence is a fault becomes rather hopeless in practice. It cannot be excluded that a fragment that has been found faulty and has been modified at some stage must be restored to its original form in a later stage because another ``deeper'' problem has surfaced the solution of which as a side effect reinstalls the mentioned fault as not faulty and its repair (once it has been applied) as a candidate fault with the original fragment as a promising candidate solution.

The idea that incorrect instruction sequences, that is instruction sequences for which effectuation failures have been noticed, can be transformed in a stepwise fashion to correct instruction sequences by means of successive identification and reparation of faults is a fact of life that is supposed to come from sustained (hypothetical) observation of (a hypothetical) instruction sequence engineering practice, taking programming practice as a source of this idea. The contrast between formal verification and testing only arises if one forgets that from the outset supporting transformational improvement engineering has been an essential objective for the application of experimentation tests and confirmation tests for finding failures and identifying faults. The question about the contrast between verification and testing can be alternatively phrased as follows: what has verification to offer if one intends to enact a stepwise transformation from an incorrect instruction sequence to a ``more correct'' one?

In program algebra \cite{BL02a} a variety of different instruction sequence notations, each with in essence the same expressive power, has been developed and transformations allow to rewrite instruction sequences into different notations. Some notations are more high level than other notations and the relation with faults seems to be the following. When translating $x$ written in a more high level notation to a more low level notation, $k$-located faults may become $l$-located faults with $l>k$. In other words, the more high level a notation is, the better the chance that an effectuation failure is caused by a 1-located fault.

\subsection{The teleological classification helps to avoid a circularity}
The necessity (or at least the usefulness) of a teleological classification of effectuations is seen as follows:  in order to see 
(check, validate, establish) that $f$ is a fault in $x$, its potential of reparation must be established by actually showing that that can be done (that is finding $f$ and designing $f^{\prime}$ and adapting $x$ accordingly). That constellation then gives rise to a confirmation test in order to establish that the repair works in the case that was the trigger for its design. Viewed as an effectuation of $f^{\prime}$ it is classified as a test because it cannot be classified as a form of use. As a test is is not performed with the intention to find a failure or to locate a fault, for the simple reason that that has been done already concerning $f$ and therefore it is not an experimentation test. Summing up: some (instruction sequence) tests are needed to certify that certain fragments constitute faults and such tests are not performed in order to find faults. This state of affairs cannot coexist with the uncompromising, but quite common, view that all ``instruction sequence testing is instruction sequence effectuation with the intention of finding faults''.

\subsection{Adequacy modulo a limited volume of faults}
An instruction sequence $x$ is adequate modulo a limited volume of faults if (i) it has a finite collection of disjoint faults, which (ii) together do not exceed 25\% of the size of $x$, and (iii) by successively repairing these faults a correct  (that is, not producing failure in any effectuation) instruction sequence $x^{\prime}$  is obtained from $x$.

Adequacy modulo a limited volume of  faults, is a quality level for an instruction sequence. It indicates that by merely repairing a limited collection of faults an adequate instruction sequence is obtained. For computer programs in practice this quality level makes sense as well, and it is often only obtained after extensive use of a program.

\section{Instruction sequence software defects}
\label{sect-defects-faults}
The term software is often used as an abbreviation of computer software which in its turn stands for computer programs and  related software, or more briefly, computer program software. That includes requirements capturing documents, specifications, text based verifications, test reports, inspection reports, reports on automated verifications, and comments, as well as historic information about a system's growth and evolution. 
I will use ``instruction sequence software'' as that form of software which is about instruction sequences and their requirements, specifications, verifications, and comments.

The phrase ``software defect'' is often used, for instance in \cite{Wagner2006} the notion of a software defect is used which extends beyond that of a program fault.%
\footnote{In \cite{Mays1990}, still using some now outdated terminology concerning errors instead of faults, defect prevention concerns all possible defects of products and processes. A systematic approach towards defect prevention is outlined, which today might perhaps preferably be called process innovation.} 
If one adheres to the mechanical understanding of fault, software defect combines faults that have mechanical significance with specification flaws which have no mechanical impact. Of course a fragment in a specification can just as much constitute a cause of a failure as a fragment of an instruction sequence. This grouping together of mechanical causes of failure with non-mechanical causes of failure justifies the use of defect as a term in addition to fault.

An instruction sequence software defect may either be an instruction sequence fault, given a specification that is left implicit, or a specification defect, which is a mismatch between the specification and ``what was really intended, even when not specified''. Forgetting about requirements one may split instruction sequence software in instruction sequences and specifications. Now it is reasonable to assume that specifications can have defects but no faults and that instruction sequence software may have defects which include both instruction sequence faults and specification defects.

\subsection{Defects versus faults}
Defects can originate in a requirements capturing phase and percolate all the way into the test generation 
process. In \cite{HewettK2009} one finds the claim that software defects migrate through a specific life-cycle, and contemplation of that life-cycle at the same time provides a useful fine structure for the debugging process.%
\footnote{Some authors (e.g. \cite{Alzamil2008}) assume that program defects can be localized, which suggests that ``program defect''  and ``program fault coincide''.}
At the lower side of the spectrum \cite{MukherjeeER2005} discusses soft-errors which are rather random unwanted bit-flips in a microprocessor, which may be radiation induced.%
\footnote{As dynamic events these are rightly called errors, however, the term error is specifically used in the case that error correction fails to remedy the original bit-flip, which itself is referred to as a fault, or rather the occurrence of a faulty bit. This terminology is rather confusing, and it would be more consistent with software engineering terminology to consider the flipped bit an error which, when left undetected or uncorrected proceeds as a failure, which may subsequently affect the input of other hardware or software. Unlike software faults the occurrence of soft-errors requires remedies that make use of duplication techniques and of error correction protocols, all ending up as features of a mircro-processor architecture. In principle one can imagine the same strategy for a program: given its specification, write several implementations of it and perform majority voting on every outcome. That strategy, however, seems to play a marginal role 
only in the literature on software faults.}

Program faults cannot exist in the absence of a specification (requirements document). It is the mismatch between behavior when put into effect and the expected behavior as specified which may be detected 
experimentally during a test and which is considered a failure for which the programmer is held responsible. The programmer's mistake was to embed a fault in the program, thus causing the failure. Indeed the phrase ``program fault''  is somehow incomplete. The specification seems to be always forgotten, if only in the terminology, and instead one might also postulate the existence of a ``specification fault'' (or rather a specification defect) in the case of an observed failure. One might think that the specification is by default more 
important than the program, thus allowing to put the blame on the program, also by default. But at the same time the specification is left implicit quite often.%
\footnote{Assuming that requirements are captured in a document which constitutes part of the software, 
a fault in the
requirements is a software fault, while it is not a program fault. In \cite{BourdonovKK2007} an error (fault, failure) which needs to be repaired by changing the requirements is called a phantom error, which in the terminology that I am using would rather suggest the presence of a requirements fault. In principle it is possible to speak of phantom faults, because phantom errors may be caused by the effectuation of program fragments that can be physically located in the program. One should never intend to repair a phantom fault, though one may not be aware that the requirements are faulty, in which case exactly that may be pursued in spite of it being headed the wrong way. It my be useful to speak of phantom failures, however. At this point one may notice a significant discrepancy between software faults and program faults, which translates into an equally significant difference between ``software testing" and ``program" testing if conceived in terms of the objectives of fault detection and fault location.}

There is a further extension that defect comprises beyond fault. It may be the case that an effectuation failure occurs concerning instruction sequence $x$  but that no fault (or $n$-located fault) within $x$ can be blamed for being its cause, and instead a complete redesign of the instruction sequence needs to be performed in order to obtain an instruction sequence $y$ that delivers fewer effectuation failures than $x$. In this case the problem with $x$ is avoidable by working with $y$ instead (not by adapting $x$), and it is not the case that a localized fault within $x$ admits a local solution of the observed effectuation failure.

If $x$ is adequate modulo a limited volume of faults precisely if  each instruction sequence defect of $x$ is a fault of $x$.

\subsection{Instruction sequence inspection and fault removal}
Assuming that an instruction sequence is constructed on the basis of a requirements specification, then inspection by human readers is likely to detect many candidate faults. Such candidate faults are instruction sequence fragments where at a closer inspection a modified fragment is judged to be more likely to work as intended. A candidate fault found during inspection need not constitute a fault in the mechanical sense.%
\footnote{In particular it can be the case that two candidate faults compensate each other's effects so that effectuating an instruction sequence containing both shows no failures. The effectuation may reveal errors, however, which do not perpetuate into failures. Then correcting the first candidate fault may constitute no less than the introduction of a fault (that is the new status, after instruction sequence modification,  of the second original candidate fault containing the compensating activity), whereas correcting the compensating fault is expected to resolve that failure again.}

In \cite{Dyer1987} it is stated that software inspection can precede testing in such a way that fewer faults are (and need to be) found, whereas formal verification can partially replace testing.

\subsection{Spontaneously generated requirements}
The intuition seems to be that even if the specification is thrown away, still some notion of program-fault survives. This seems to have happened in the Y2K case where defects in the light of the Y2K transition were considered program faults in spite of these programs having been conceived without any ambition to survive Y2K without significant modification. By having been used for many years the programs had created their own expectation of functionality in the minds of a not completely informed user community, in spite of distinctly different objectives of the program's designers decades before. It is against these  spontaneously generated expectations
that the programs were considered to be at fault. Although that has been unfair towards the owners and the designers of these programs, the very terminology of program faults already carries with it the priority that programs are expected to  take over their specifications.

Perhaps any artifact class, after having been conceived and after having been inhabited with a variety of artifacts, will create an automatic intuition or an implicit concept of fault for that class, because of an overwhelming incentive to separate the artifact class from the specific design histories of its members. If a tool used by a surgeon has not been thoroughly cleaned and safely put away in an orderly fashion until the 
moment of use it cannot be sterile. 
If that tool is used in spite of this fact one will say that the tool was 
``dirty'', with the implication that somehow the tool was faulty. But of course it was not, the process was at fault, with the dirty tool as evidence of that fact but not as a cause. The tool being dirty is not a property of the tool. Similarly many so-called program faults can't be considered to feature properties of the programs in which these faults are claimed to reside.

\subsection{Unstable requirements}
In \cite{Schneidewind1987} it is claimed that requirements need not be finished before the start of an implementation. According to that view requirements can also emanate from the use experience and requirements capture does not end in the requirements phase. These arguments are used in \cite{Schneidewind1987} in order to prove the need of systematic software maintenance methods. For the notion of a program fault it means no less than that a faultless program may become faulty. Complementary to the terminology of \cite{BourdonovKK2007} it is then plausible to speak of phantom correctness of a program if the invalidity of a set of requirements has not yet been discovered, so that corresponding failures were missed and faults left undetected and unlocated.

\subsection{Persistent faults}
In \cite{Glass1981} the notion of a persistent software error is proposed, which I presume in more recent terminology should be referred to as a persistent software fault. As defined in \cite{Glass1981} a persistent fault is a fault that survives until the stage of use. Persistent software faults seem to exist of two kinds only:  persistent program faults and persistent requirements faults.%
\footnote{The terminology has many variations, for instance as recent as 2000 in  \cite{Whittaker2000}  one may find ``(a)ssuming the bugs users report occur in a software product that really is in error, ...''. I assume that bug equals fault and that users report failures rather than bugs, the product being faulty rather than being in error.  In terms of the terminology that I am using this statement would translate into: ``(a)ssuming that the failures users report in a program are non-phantom, and assuming that a permanent fault causes such a failure, ...''. It is remarkable that a relatively recent IEEE practice tutorial completely 
ignores the terminology of fault, error and failure, while suggesting that research on testing is too academic.}
The notion of a persistent instruction sequence fault is problematic because instruction sequences are a model of programs for which the term use cannot be applied in a fully realistic sense.

By definition testing cannot detect persistent faults but it can prevent a fault from becoming persistent. The same holds for inspection and for verification. Persistency is not a statical property of a fault definable in terms of the instruction sequence and its specification or requirements. Instead  persistency of a fault must be seen in the context of the entire life-cycle, at least until the moment of detection of the failure from which the fault's existence is inferred, (if that ever happens of course).

Nevertheless as \cite{Glass1981} indicates one may perform statistics on persistent faults given a number of projects and that leads to the conclusion that such faults correlate with a programmer's (nowadays designer's) underestimation of the algorithmic complexity of the task that had to be implemented.

\subsection{Taking the decision to use an instruction sequence}
At some stage, if (the effectuation of) an instruction sequence $x$ is ever used, it may be the 
case that a decision to do so has been  taken in preparation of usage. Following
\cite{Bergstra2011d} the mere fact that $x$ is effectuated by way of use does not imply that a 
decision to do so was taken. 

In \cite{Bergstra2011d} a detailed definition of ``decision''  has been elaborated according to which 
significant constraints are imposed on any decision taking thread.
A natural precondition for the decision to use instruction sequence $x$ is that the collection of persistent software defects
concerning $x$ is judged to be sufficiently small in number and impact so that the risks emerging from these persistent software defects can and will be accepted. Acceptance testing is often suggested as a major method to arrive at this judgement, which combines a positive judgement about the  absence of problematic defects in requirements $R$ (including specifications) with a judgement concerning the absence of instruction sequence faults, given $R$.

In order to validate the requirements $R$ that $x$ is supposed to obey, one may apply experimental tests of $x$ (so that $x$ plays the role of a prototype). Doing so can only be helpful for validation of $R$ if the conformance of $x$ with $R$~has been investigated and sufficiently confirmed. That may lead to a two-stage testing process as follows:
\begin{enumerate}
\item ({\em Verification of $x$ w.r.t. $R$.}) Investigating the presence of failures when effectuating $x$ by way of experimental tests, given $R$, while accepting that some failures that are found, and for which a cause in the form of a fault is found and subsequently is repaired, may in the second  stage prove to have been phantom failures, and
\item ({\em Validation of $R$ with help of $x$.}) Investigating the validity of $R$, while assuming the correctness of $x$ w.r.t. $R$. In this stage $x$ is used as a prototype implementation of $R$. If modifications of $r$ are proposed, these modifications will give rise to a redesign of $x$ after which the two stage process can be repeated. This repetition proceeds until $R$ has stabilized.
\end{enumerate}

The higher the decision taker's confidence is in the requirements $R$ that have been laid down, thus accentuating that the risk of a persistent requirements defect is low, the more important it is that persistent program faults are few and of marginal importance. 

Acceptance testing of instruction sequences must be helpful in producing an assessment of  the risks caused by persistent defects. Achieving that assessment is more easily said than done and systematic literature about testing as a help for deciding about use is rather hard to find.

\section{Four views on instruction sequence faults}
In the previous sections two views of instruction sequence faults have been contrasted: the logical view, or if one prefers the mathematical view, which has been advocated in \cite{Laski1997} and the empirical (that is confirmation test based) view, for which I have not found satisfactory sources in the literature on software faults, in spite of the fact that it seems to be a majority view. 

I have rejected the logical/mathematical view because of its implausibility, thus giving way to the suggestion that the empirical view is more plausible, notwithstanding the fact that the empirical view makes the concept of fault dependent on the concept of test. This is problematic if one assumes that the concept of test in its turn is explained in terms of faults (more specifically the objective to discover failures as indications for faults). By distinguishing confirmation tests from experimentation tests a circularity can be avoided. 

Both the the logical/mathematical view and the empirical view are special cases of the following more general view ``instruction sequence faults as avoidable causes of effectuation failures". Two radically different views on instruction sequence faults are possible, however: ``instruction sequence faults as avoidable causes of negative expert judgement", and ``instruction sequence faults as implicit causes of incorrectness''. Experts split in two types: instruction sequence product experts who make assessment of particular instruction sequences viewing them as products subject to quality control, and process experts who make assessments concerning the entire production process rather than of specific outcomes of such a process. Together this leads to four conceptions of an instruction sequence fault.

These four views seem to be competing rather than complementary. I hold that different schools of thought can exist in which the term fault plays these different roles. These different schools of thought can support entire engineering philosophies and their participants need not even know of the existence of alternative views.

\subsection{Mechanical view: instruction sequence faults as avoidable and localized causes of effectuation failures}
In this view one considers an effectuation failure an observation from which by means of appropriate diagnostic methods a cause is derived. The cause usually consists of an instruction sequence fragment (playing the role of the fault), together with a suggested replacement of that fragment (playing the role of the fix of the fault), so that, (i) had the replacement been installed instead of the (faulty) fragment the failure had been avoided, and moreover (ii) the replacement has not given rise to any potential new failure which had been observed to be absent until the moment in the engineering phase that the replacement was made (that is regression tests have been made),%
\footnote{A regression test is an experimentation test intended to demonstrate that an effectuation that was valid (not failing) before the replacement for a fault by its fix remains valid after the fix is applied. Thus until the complete series of regression tests has been performed fault and fix have the status of candidate fault and candidate fix. Regression testing cannot prevent that in a later stage the fix must unfortunately be considered a fault.}
or (iii) there is adequate evidence that replacing the fault by its fix will not introduce any new failures.

This view easily translates from instruction sequences to programs.
Many papers on program testing are written with this mechanism in mind. It may even constitute the dominant view on faults in the program testing community.%
\footnote{Though it seems not to be the dominant view on program faults in programming education. There the 
``instruction sequence faults as avoidable causes of negative expert judgement" view seems to be far stronger, even if testing is advocated to find faults of that kind.}
The mechanistic view of faults has two main flavors, with two flavors in between:
\begin{description}
\item{\em Logical view of fault mechanics.} The concept of an instruction sequence fault requires a logical or mathematical definition. This definition presupposes a proof system for instruction sequence correctness and a formal operational semantics.%
\footnote{The view on faults of \cite{Laski1997} falls under this category. Remarkably these views had no echo in the world of formal methods and program verification, probably because there the incentive to explain the absence of a correctness proof via the occurrence of a formally defined concept of fault is nonexistent. Formal methods people seem to prefer the
 ``instruction sequence faults as implicit causes of incorrecteness''  view of faults.}

\item{\em Confirmation test supported idealized logical view of fault mechanics.} In this view the idea of a logical definition of fault is used in the background. In other words, although fault is conceived of as a concept
for which a formal and logical definition exists, the existence of that kind of definition only serves as a hypothesis, thus rendering the notion of a fault a hypothesized one as well. Confirmation test based notions are used to obtain approximations of the hypothesized notion, while the method of approximation is left implicit. In this approach false positives must be expected, while applying a repair may not remove a fault.

\item{\em Formalized (and confirmation test supported)  confirmation test based view of fault mechanics.} In this view the confirmation test based view is ``formalized''. That means in particular that instead of an idealized repair criterion one uses a formalized version of the requirement that regression testing confirms that a candidate repair introduces no new faults.  In addition confirmation test support is permitted which amounts to the following:  the proof that an effectuation leads to some result thereby witnessing (or refuting) a particular failure is (or may be) replaced by giving evidence and an effectuation 
may be taken for that evidence even if no compliance with a  formal operational semantics has been established.%
\footnote{An instruction sequence loaded in a machine and capable of influencing its behavior 
constitutes a physical object, 
it cannot be a mere mathematical object. Such an instruction sequence containing a fault can only be defined in informal though conceivably rigorous ways.

At the opposite end of the spectrum of meanings of ``instruction sequence''  an instruction sequence may be given as a syntactic object meant to represent a mathematical term in a way most useful for performing mathematical transformations and reasoning steps. Of course the mathematical reasoning process operates by means of the manipulation of physical representations just as well, and at some stage the notion of a proof becomes dependent on the uncertainties of the physical world. The latter uncertainties, however, are independent of the subject of mathematical analysis and are mainly dependent on the size and complexity of the structures representing various mathematical entities that are being manipulated.

Along the lines proposed in \cite{Laski1997}, and given a mathematical presentation of the 
operational semantics
of an instruction sequence notation, a correctness proof system can be developed that matches the requirements of \cite{Laski1997} in such a way  that the combination of these two (that is operational semantics and proof system) can be used to provide a satisfactory logical definition of an effectuation failure and based upon that of an instruction sequence fault which causes the failure. This formalized view of faults depends on the precise match between the formal semantics and the machine behavior, an issue which may be quite hard to determine, and in particular an issue which itself may depend on making inferences from experimental data, most likely coming from a variety of exploration tests, specifically gathered to validate the assertion that the operational semantics can be used as a theory for predicting behavior. I expext that, during a systematic empirical process designed to resolve these matters, already before the volume of experimental data that has been gathered is sufficiently large and robust to allow to infer the existence of a perfect match between formal semantics and empirical semantics, notions of (effectuation) failure and (instruction sequence) fault will have emerged. These notions will be needed for a systematic organization of the experimental process meant to guarantee that effectuations of instruction sequences are in conformance with the given semantics of the instruction sequence notation.

If faults are defined formally, the presence of a specific fault $f$ in an instruction sequence $x$ must be considered a mathematical fact requiring a mathematical proof. In that proof the effectuation of $x$ on a machine and the observation of a failure which is consistent with the predictions made by the formal semantic model of $x$ may be helpful, but its status is in need of careful analysis before it can be admitted as a component of a mathematical proof.}

\item{\em Confirmation test based view of fault mechanics.} The concept of instruction sequence fault is based on an analysis of experiment and observation for instruction sequence effectuation on a specific computing platform.

\item{\em Confirmation test based view of fault mechanics combined with expert judgement supported repair assessment.} The concept of instruction sequence fault is based on an analysis of experiment and observation for instruction sequence effectuation on a specific computing platform. The assessment of what constitutes a repair combines regression testing with expert judgement for bridging the gap between the limited scope of the regression tests and the idealized criterion for a candidate repair.
\end{description}

I hold that the confirmation test based view of fault mechanics is closest to what most software engineers have in mind. But it is not the whole picture, the occurrence of false positives in fault identification (due to false positives in candidate  repair assessments) and the potential replacements of faults by new faults, is remedied by the engineer's expert judgement that the proposed candidate repair will work ``in general'' and that regression testing merely confirms that judgement but in no way replaces his or her expert opinion. This leads to the fifth view of mechanical faults as a most plausible approximation of current ``practice''.

Predicting preferences between these different views presents additional complications. I mention some aspects:
\begin{description}
\item{\em Implausibility of the idealized logical view of fault mechanics.}
I conclude that the logical/mathematical view of faults, and also the confirmation test supported idealized logical view, is self-defeating in practice because of its logical complexity. 
\item{\em Plausibility of the confirmation test supported logical view of fault mechanics.} I hold that for current practitioners of software engineering a mixture between a  logical view on fault mechanics and a confirmation test based view of fault mechanics is to be expected. I expect that when asked to provide a definition of an instruction sequence fault software engineers will converge to expressing the need for a logical definition of fault without going at lengths to provide one, and at the same time they will assume that effectuations are a valid way to collect information that can be imported into a logical definition to such an extent that the logical formality of the approach is essentially lost.
\item{\em Complexity of the causal analysis.} What makes the causal story of faults especially difficult is that large instruction sequences may contain many faults. Then one needs some kind of correctness metric to express that fixing a single fault constitutes a step forward, even if the result is an instruction sequence that still contains many faults. Formalizing this complex matter has only been partially successful in \cite{Laski1997} where a lot of work is already needed to determine faults under the assumption that there is just one fault present. Dealing with a multiplicity of faults imposes different requirements on the different views of mechanical faults, and the ease of incorporating fault multiplicity in a particular view may influence one's acceptance of that view.
\end{description}

\subsubsection{Numerical boundaries for faults and fixes.}
When working with instruction sequences it is reasonable to require that a fault may not be longer than say $10 \%$ of the entire instruction sequence, with their fix deviating in length max $50\%$ from the length fault. Such numerical bounds are quite arbitrary, but somehow needed in order to obtain a clear definition of ``fault''.%
\footnote{Alternatively on may entertain ``instruction sequence fault'' as a fuzzy notion based on a concept of faultiness (being faulty) which depends on a combination of numerical paramemeters.}
One may hypothetically run all possible tests, and derive from that all possible failures and subsequently all faults in a given instruction sequence. This gives a clue on how to count the number of faults at least in principle, even if that number is never established. 

Further it may be assumed that the total sum of lengths of faults detected during an instruction sequence development process is not more than say $25\%$ of the length of the initial instruction sequence. If a larger fault is needed, the corresponding repair is labeled a redesign of the instruction sequence and all quality control activity, including various tests, needs to be redone from scratch.

If the need to redesign arises one may assume that an infinite number of faults has been noticed. Thus an instruction sequence cannot contain more faults than its length, unless the number is infinite, which indicates that ordinary fault removal has not worked after all.

Of course these bounds are arbitrary to a large extent. But there seems to be no formal concept of an instruction sequence fault in the absence of such bounds. That holds for the empirical approach to avoidable causes of failures as well.

\subsubsection{Strictly logical mechanical fault, test confirmed mechanical fault, and mixed view mechanical fault.}
If communication with persons holding other views on faults is to be maintained it may be helpful to speak of mechanical faults, whenever avoidable causes of effectuation failures are meant. Different views on mechanical faults may be reflected in different phrases: strictly logical mechanical fault in the case that a fault is identified without making use of confirmation tests, test confirmed mechanical faults in the case that all relevant information concerning the outcome of effectuation results from confirmation tests and experimentation tests. If logic and confirmation tests are combined to find evidence for a fault it may be helpful to speak of a mixed view mechanical fault.

These specialized terms for mechanical faults may be in need of alternatives. For instance instead of a logical mechanical fault one may speak of a formally confirmed mechanical fault. Instead of a mixed view mechanical fault one may prefer to speak of a mechanical fault with hybrid (combining confirmation tests and  formal deduction) confirmation. The term mechanical fault admits alternatives too, for instance: failure causing fault, failure related fault, failure based fault, 
failure relevant fault, reliability obstacle, compliance obstacle.

In the setting of mechanical faults the relation between cause (fault) and effect (failure) seems to be obvious, but it is not. Observation of a failure may induce that a fragment of an instruction sequence is labeled as faulty, and so the causal 
relationship between fault and failure must be considered at different levels. Only in a context where the instruction sequence as a whole is considered a cause of the behavior observed during an effectuation, it makes sense to view a particular fragment of the instruction sequence as a cause for a particular deviation of that behavior from a somehow given intended behavior.

\subsection{Product authority view: instruction sequence faults as avoidable and localized causes of negative judgements by  instruction sequence product experts} 
In this description I will phrase in terms of instruction sequences what I consider to be a plausible conception of program faults. This is done in order to obtain a coherent picture for the four conceptions of faults. Unfortunately thinking is terms of instruction sequences is not particularly helpful for understanding the product authority view.

It may be imagined as a thought experiment that whoever constructs an instruction sequence may imagine the additional presence of an expert consultant on instruction sequencing. This (hypothetical) expert knows when fragments of instruction sequences are faulty. The expert also knows which requirement specifications are acceptable and should be adhered to and which specifications are problematic and in need for further scrutiny before being taken literally as a  guide for implementation design. Instruction sequence faults are understood as fragments of instruction sequences in exactly the same way as in the mechanical conception of faults. Thus faults have a size and may be distributed over several locations. Further different faults are disjoint and faults can be counted.

In this hypothetical world, instruction sequencing experts are equipped with a  community 
competence on fault detection and fault fixing 
using the terminology of \cite{BDV2011b}. Programming style is more important for  an expert than having a theory at hand for predicting system behavior. Labeling a fragment as a fault in no way implies any obligation for the expert  to indicate for what inputs an experimental test is expected to uncover a failure. Labeling a fragment as faulty by the expert is reasonable even if (s)he knows that no effectuation failure is to be expected from the faults existence.

\subsubsection{Defects in the product authority view.}
In the product authority view an instruction sequence defect may of course be any fault  in the sense of the product 
authority view. Thus the notion of an instruction sequence (software) defect depends on the underlying view of fault.%
\footnote{A mere violation of coding standards is not considered an instruction sequence defect in the mechanical perspective on faults. The mechanical view needs ``compliance with instruction sequence coding standards (if any are present)'' as an additional quality criterion on top of the absence of instruction sequence software defects. In asserting these particularities about the meaning of (instruction sequence) defects the freedom to determine the meaning of terminology concerning instruction sequences is used. My
primary objective is to develop a clear an consistent terminology concerning instruction sequences and not to reflect
some existing judgement concerning the question whether or not violations of coding standards constitute software defects.}

\subsubsection{Different view or different meaning assignment for identical terms.}
In the product authority view of faults the slogan ``testing can prove the existence of faults but not their absence'' is wrong,
unless testing is assumed to include inspection (which some people do), for reasons differing from those that invalidate the same assertion in the mechanical view.

It is easy to explain the contrast between the mechanical view and the product authority view in terms of a combination of mere differences in meaning combined with a differences in emphasis.

Indeed a simple way to understand the product authority view from the mechanical view is to conceive of it as a a view in which: (i)  instruction sequence defects are the key notion,  (ii)  failures occur as a secondary notion,  (iii) there is no wish to understand failures as caused by mechanical faults, (iv)  the ``hierarchy'' of mistake, fault, error and failure plays no role, and, (v) the term fault is used where a proponent of the mechanical view would prefer the term defect (after he or she has admitted that violations of coding standards and other triggers of negative product expert judgement are defects as well).

If defect is understood to be more inclusive than a proponent of the mechanical view might need to explain his/her own views, both views can be integrated if the proponents of the product authority view adapt their their usage of the terms fault and defect and only speak in terms of defects, and not in terms of mistakes, faults, and errors.

\subsubsection{Architectural considerations.}
Besides being small in terms of textual size, candidate faults and corresponding candidate repairs should be small in architectural terms as well. An architecture is determined as an abstraction of an instruction sequence, either in advance or by way of reverse engineering, and  repairing faults must leave the architecture unchanged. Having agreed upon an architecture that explains the structure of an instruction sequence, changing that architecture amounts to redesign and is not a matter of fault repair anymore. Appreciation of architectural patterns within an instruction sequence is primarily a matter of product expert competence. For instance a product authority may insist that an instruction sequence is designed in such a way that its architecture is recoverable with relatively ease.

\subsubsection{Community confirmed competence based view of fault.}
In the community competence view of a fault, an instruction sequence fault is a fragment of an instruction sequence which is bound to be considered faulty by competent peers (individuals whose competence concerning instruction sequence production and instruction sequences fault assessment is confirmed by a suitable community).
Following \cite{BDV2011b} it is assumed that a person or a team developing an
 instruction sequence imagines the presence of a team of peers who are considered very strong and  experienced in instruction sequence production. A fault is precisely what these (possibly hypothetical)  peers would qualify as a fault.
 Because in many circumstances a fault impacts the behavior of an instruction sequence effectuation, the observation of failures in that behavior can provide a pointer towards finding that fault.
 
 \subsubsection{Natural perspective in the context of instruction sequencing education.}
 For most programming exercises which are used when teaching person $P$ how to design and write an instruction sequence for a specified computing platform, given a requirements specification, the hypothesis that an expert instruction sequence constructor  exists who may distinguish adequate from inadequate (that is faulty) fragments in an instruction sequence is reasonable. A teacher or a consultant supporting the teacher may well fit that role. Finding faults is doable by means of inspections as well as by means of effectuations classifiable as experimentation tests. Having performed adequate experimentation tests is likely to be part of the advice that is expected from an expert computer programmer. That is, the expert expects to be consulted only after a systematic amount of experimentation testing has 
been performed and after observed failures have been traced back to causally related faults for which improvements have been found and applied to the instruction sequence under construction. 

The hypothetical existence (in $P$'s mind) of an expert able to decide about program fragments being faulty is an inductive generalization of the learning context for $P$. So $P$ considers a program fragment faulty if it is plausible that an expert instruction sequence designer would consider that fragment to be at fault. This criterion abstains from any prediction of a failure being caused by the fault. Being faulty already begins with a fragment not conforming to coding standards which the expert advocates. 

\subsubsection{Violations rather than faults.}
If $P$ intends to avoid confusion when communicating with persons having different views on faults, $P$ may 
refer to the faults
indicated by product expert $E$ as ``violations of art'' (including style, method, coding standards, 
rules of architectural design), or simply violations. Some violations are faults in a mechanical understanding and some may not qualify as faults in a mechanistic view.

\subsubsection{The product authority view enters the use of the confirmation test based mechanical view.}
Given the fact that a confirmation test based view of mechanical faults is vulnerable for the occurrence of false positives in the judgement of candidate repairs because of the limitations of regression testing, a candidate repair must be assessed (for its potential of effecting an improvement) by an expert judgement as well as by means of mere regression testing in order to create the confidence that enacting the repair constitutes a step forward towards a fault free instruction sequence indeed.

 \subsection{Incorrectness substitute view: instruction sequence faults as implicit and hypothetical causes of incorrectness}
 Although formal verification based software engineering claims to be helpful in the construction of so-called bug-free programs, the notion of a bug (that is a fault) receives little attention. If a program cannot be verified it is the best option to modify the program until that can be done. There is no need to think in terms of faults that cause incorrect behavior, it suffices to think in terms of modifications of the program that allow verification.
 
If a program, or an instruction sequence, exhibits incorrect behavior when effectuated it is common practice to call it faulty, and that seems to imply that it contains faults. These faults have a hypothetical status as they will (or may) not be physically spotted in the process of instruction sequence improvement and proof engineering aimed at achieving provable correctness. Nevertheless for those who think in terms of correctness speaking of faults may be a way to avoid speaking of incorrectness which has a less practical connotation for some part of the audience, in particular for those who see ``program correctness'' as a largely academic exercise. Without a clear definition of fault the identification of incorrectness with the presence of one or more faults cannot be taken for granted. In fact it is rather implausible that such an identification can be established.

For workers in formal methods the simplest way to explain the concept of instruction sequence correctness to those not thinking in terms of correctness  is to phrase it as the absence of faults. Speaking of the presence of faults can be used as a way to find a substitute for  the notion of instruction sequence incorrectness.

This view is where the slogan ``testing can prove the presence of faults but not the absence of faults'' resides.
The incorrectness substitute  view is a natural perception for those who favor formal specifications of (intended) instruction sequence behavior and corresponding formal verifications. Instruction sequencing engineers with such preferences will use the term fault in order to avoid the use of a formalist language that is likely to sound impractical to uninitiated colleagues. At the same time the technical notion of  fault will be of little relevance to their work. 

If a notion of fault  is used at all when working with this perspective it is understood as a cause of the failure to obtain a proof for one of the given requirements rather than as the cause of an effectuation failure. In the place of a regression test one finds the task to redo proofs of requirements that had been established already before  the modification was applied. In this view the term fault may well be replaced by the phrase ``proof obstacle'', thus reducing the risk of confusion with other uses of fault.

\subsection{Process authority view: instruction sequence faults as hypothetical justifications of avoidable negative judgements by instruction 
sequence process experts}
Just as an expert on instruction sequence structure may entertain opinions concerning what instruction sequence notations, and given such a notation what instruction sequences and what instruction sequence fragments are to be appreciated, thus giving rise to the community confirmed competence view on faults, experts on instruction sequence construction may hold that testing (however defined) is an essential feature of sound instruction sequence engineering. Such engineering oriented experts (also termed instruction sequence process experts) may hold that testing must be performed in such a way that it can discover failures and that by definition faults are those causes of failures found in that way which can be repaired by way of modifying the instruction sequence under test (playing the role of PuFA). 

In this view on faults an engineering process must contain testing phases and faults are found as a consequence of testing.  Finding and removing the faults is used as a way to justify some of the operational rules enforced or advised  by 
the instruction sequence process expert.%
\footnote{In other words faults, which are not actually identified, but rather expected on statistical grounds,  and for that reason may be thought of as  hypothetical, are used as a justification, for negative judgements concerning steps taken in the production process.}
The expert may hand over the testing activity as well as any assessment of what constitutes a fault to a subordinate engineer, allowing any of the three conceptions of faults mentioned above, or any other conception of fault the engineer in charge of testing may put forward, even without feeling the need of being informed about which conception of fault is assumed and why that is done.%
\footnote{The instruction sequence process expert may hold that no specific conception of instruction sequence fault is needed because each conception has its own virtues. In other words, this view allows the expert to abstract from the concept of a fault, without throwing it away. The coherence of this fourth view seems to be problematic but its occurrence in practice seems to be an empirical fact.}

Whether or not an engineering process involves adequate testing is a matter of expert judgement rather than of a match between an activity and a definition of testing as such. In some cases it may suffice for the expert if certain effectuations are called testing to assert that testing has been performed. Instruction sequence modifications performed as a consequence of these ``tests'', however causally related to observations, will then be considered fault reparations, and the original fragments are considered faulty.

So in this fourth view, a confirmed community competence about instruction sequence engineering leads to expert judgements on testing and an informal argument views faults as correlated to tests. In this view tests are not defined in terms of faults but rather the other way around. Using benchmarks found from comparable engineering projects the 
instruction sequence engineering expert may for instance claim that testing should preferably make up say 50\% 
of the engineering effort, and the expert may stop testing once that amount of time has been spent to activities classified as testing, under the assumption that this warrants sufficient scrutiny. A potential weakness of the fourth view is that the expert may be unable to explain when testing can stop, unable to produce an independent definition of fault, and unable to produce an independent definition of failure. That lack of clarity may lead to the risk that validation and verification are confused and both of these aspects are left in a defective status when instruction sequence usage begins.

\subsubsection{Process quality violations.} A process expert may see to it that certain violations of an adequate engineering process are avoided. Here are some examples of violations that matter within the context of this paper.
\begin{description}
\item{\em Insufficient instruction sequence understanding available when designing a candidate repair.}%
\footnote{Instruction sequence understanding is the (hypothetical) instruction sequence counterpart of program understanding.}
This maters a lot because even in the confirmation test based mechanical view of faults it must be assumed that a candidate repair is design in a context of an adequate understanding of the design rationale of the entire instruction sequence. In the absence of that understanding the risk of false positives after regression testing would be too high.
\item{\em Insufficient instruction sequence engineering competence.}
A violation of process quality may occur when engineers are at work who have been insufficiently educated and trained for the job that has to be accomplished.
\item{\em Excessive focus on coverage testing.}
If testing focuses on coverage only, functional testing may be neglected too much. 
\item{\em Unconvincing solution of the oracle problem.} 
For functional testing the process expert needs to assure that the oracle problem has been solved in a satisfactory fashion. That judgement needs to be made for each individual production process because there is no general method available for producing test oracles.
\end{description}

In each of these cases a process authority may wish to explain the rationale of the requirements on process quality in terms of avoiding the occurrence of faults in the product under construction.
\section{Compatibility of the four views and conclusions}
Each of these four views is coherent. A person need not choose one, rather it is plausible that someone's view fluctuates given the context of work. I will now contemplate a person $P$  who selects a view (on faults) depending on accidental circumstances, not being strongly committed to the viewpoint. Here are some possibilities for a context and a plausible viewpoint given that context.

\begin{itemize}
\item
In the presence of a self-confident and influential expert $E$ who claims authoritative knowledge of fault versus non-fault regarding instruction sequences, $P$ may choose for the product authority  view. $P$ will assume that $E$ has acquired a community confirmed competence regarding instruction sequence faults  which justifies $P$ in assigning to $E$ the relevant authority. 
\item
In the absence of an expert $E$ and in the presence of sufficient and trustworthy mathematical information (for $P$ and relevant colleagues of $P$) about the semantics of the instruction sequence notation used for a project, $P$ may think in terms of instruction sequence correctness and incorrectness. $P$ may speak of absence or presence of faults with those colleagues who prefer not to think in terms of correctness and incorrectness, even if $P$ has no intention to identify and locate any fragments of $x$ that may be considered faults. $P$ may well be interested in finding violations according to some instruction sequencing methodology, because these violations may correlate with verification obstacles, even if they don't constitute mechanical faults.
\item
In a context where the process authority view is leading a team member may have any particular view on faults. Different team members may entertain different views of instruction sequence faults at the same time, while still being compliant with the guidelines imposed by the process authority.
\item
In the case that $P$ has no insight in the precise meaning of instruction sequences when being effectuated on the computing platform at hand, $P$ (and $P$'s colleagues) may work with experimental methods to get a feeling of how instruction sequence processing works, and by doing so $P$ will implicitly or even explicitly adhere to a confirmation test based mechanical view of the concept of fault. By qualifying faults as mechanical faults, and if needed as strictly logical mechanical faults, as test confirmed mechanical faults, or as mixed view mechanical faults, $P$ can avoid confusion concerning the intended meaning of the term ``fault''.
\end{itemize}

\subsection{Philosophical divergence and practical convergence}
Although these four different stories of fault are incompatible in philosophical terms, each giving a different conception of fault, there is some form of convergence to be expected. For many ordinary instruction sequence construction tasks asking for the building of systems that are similar to already existing one's it is conceivable that if at some stage (before the first round of fault finding has been started) some prototype has been built, there will be a significant correspondence between the candidate faults and fixes put forward by representatives of each of these views. Analyzing this convergence in practice might lead to a unified view of instruction sequence faults in spite of the significant differences of approach.

\subsection{Weighted combination of seemingly inconsistent views}
Yet another way to understand how to integrate the various conceptions of faults is simply to accept that a combination of seemingly inconsistent views may simultaneously make up a coherent picture of the concept of an instruction sequence fault. Using a weighted combination of views an engineer may arrive at a reasonably precise picture of his or her own view of the matter depending on the engineer's role in the process and depending on the engineer's view of that role.

More specifically, a mechanical view on faults may be complemented with a product authority view in order to explain away the gap between regression testing and the idealized repair criterion. The mechanical view itself may alternate between a confirmation based view and a formalized confirmation test basted view, it may allow for various forms of confirmation test and experimentation test for supporting a formalized view. The holder of a confirmation test based mechanical view may at the same hold a logical view based on the idealized repair criterion as a hypothetical basis which unfortunately is approximated by means of regression testing for pragmatic reasons. That is a case of dealing with an instance of bounded rationality. The incorrectness substitute view may be used as a communicative option that comes in handy independently of one's underlying conception of fault. The process authority view may be used as an assumption on how the external world looks at an instruction sequence development process, in such a way that room is left for different views on faults.

\subsection{Unification of the four views is trivial, and problematic}
If product authorities speak of defects rather than faults, only speaking of faults when mechanical faults are meant, the discrepancy between the mechanical view and the product authority view disappears. A similar argument applies to the incorrectness substitute view. Incorrectness certainly indicates the existence of defects. An engineer who intends to avoid using the language of correctness and who prefers to speak of the absence of faults creates a consistent picture at once if he or she insists on speaking of defects instead. In the same way the process authority view can be supported by not speaking of faults if defects are meant.

A unification of the four views requires that besides mistake, fault, error and failure, which are needed to understand the mechanical picture, also defect and violation are distinguished as key terms, where violations may be against formal or informal product design rules (product violation), as well as against formal or informal process adequacy rules (process violation).

A difficulty with this unification proposal is that it seems to turn both the product authority view and the process authority view into peripheral considerations in a setting dominated by the mechanical view. That is potentially problematic for two reasons at least. First of all product authority and process authority have become central ingredients of software quality engineering, and secondly one can imagine a future development of software engineering in which the mechanical view of faults loses its central importance. Indeed mechanical faults exist only because of engineer's inability to write programs that implement their specifications correctly. That inability need not persist.

\subsection{Conclusions}
\label{sect-conclusions}
Four coherent views on the concept of an instruction sequence fault are described. 
It is suggested that a flexible instruction sequence engineer $P$  may switch from one view to another view depending on the instruction sequence construction or maintenance project in which $P$ is participating. For each of these four views a context can be imagined where $P$ as well as $P$'s co-workers fully endorse that particular view and dismiss the other views as either intrinsically flawed or marginal and irrelevant.

Each of the four views on faults induces another perspective on testing. Rather than asserting that the concept of testing is dependent on the concept of fault, one's view of testing depends on one's view of fault. The mechanical view of faults  has further refinements in some of which the resulting concept of fault is itself dependent on the concept of testing, and in particular on so-called confirmation testing.

This story is phrased in terms of instruction sequences instead  of programs in order to reduce the risk for being in contradiction with literature on program faults. The remarkable size of that literature makes it rather difficult to find out to what extent such risks are realistic. An inductive generalization from the limited case of instruction sequences to a more general setting of programs or software may be performed by any reader who feels sufficiently convinced by the arguments given to make that step. 

Making that kind of inductive inference step in order to find an application area concerning program faults and program testing has not been an objective in this work.  Instead an objective has been to make some progress in the direction of developing a definition of instruction sequence testing as a follow-up of \cite{Bergstra2012a}. Highlighting  the dependence of mechanical faults which are not formally confirmed on confirmation tests is the main contribution of this work to that objective.

\bibliographystyle{spmpsci}
\bibliography{TA}

\end{document}